# A New Boron-10 Delivery Agent for Boron Neutron Capture Therapy: Fluorescent Boron-10 Embedded Nanodiamonds

Bo-Rong Lin[1], Srinivasu Kunuku[2], Chien-Hsu Chen[2], Tzung-Yuang Chen[3], Tung-Yuan Hsiao[2], Yu-Jen Chang[4], Li-Chuan Liao[4], Huan Niu[2*] and Chien-Ping Lee[1]

[1] *Department of Electronics Engineering and Institute of Electronics, National Chiao Tung University, Hsinchu, Taiwan*
[2]*Accelerator Laboratory, Nuclear Science and Technology Development Center, National Tsing Hua University, Hsinchu, Taiwan*
[3]*Health Physics Division, Nuclear Science and Technology Development Center, National Tsing Hua University, Hsinchu, Taiwan*
[4]*Bioresource Collection and Research Center, Food Industry Research and Development Institute, Hsinchu, Taiwan*

Electronic mail: hniu@mx.nthu.edu.tw

Boron neutron capture therapy is a powerful anti-cancer treatment, the success of which depends heavily on the boron delivery agent. Enabling the real-time tracing of delivery agents as they move through the body is crucial to the further development of boron neutron therapy. In this study, we fabricate highly bio-compatible boron-10 embedded nanodiamonds using physical ion implantation in conjunction with a two-step annealing process. The red fluorescence of the nanodiamonds allows their use in fluorescence microscopy and *in vivo* imaging systems, thereby making it possible to conduct tracking in real time. The proposed fluorescent boron-10 embedded nanodiamonds, combining optical visibility and boron-10 transport capability, are a promising boron delivery agent suitable for a wide range of biomedical applications.

Radiation therapy[1], chemotherapy[2], and surgery[3] are the three most common approaches to cancer treatment. Radiation therapy has undergone considerable advances in recent decades, particularly in the fields of photon therapy[4] and particle therapy, involving protons[5], neutrons[6], and heavy particles[7]. Boron neutron capture therapy (BNCT) involves the use of heavy particles as an alternative to conventional radiation therapy for the treatment of tenacious cancers, such as those of the head and neck cancer[8-10] and glioblastoma multiforme[11, 12] (GBM).

BNCT is based on the nuclear capture and fission reactions that occur when a non-radioactive boron-10 absorbs a neutron[13, 14]. This form of nuclear fission produces high-energy alpha particles and high-energy lithium-7, both of which induce ionization in the immediate vicinity of the reaction. The range of ionization is 4–9 μm, which is less than the diameter of a tumor cells. After the boron-10 ions are internalized by tumor cells, high-energy particles kill the tumor cells by damaging their DNA or organelles[14-16].

Many factors can affect the outcomes of BNCT treatment, particularly the delivery agents used for the transport of boron-10 to tumor tissue. Boron delivery agents must be low in toxicity[17]. They must permit the retention of boron-10 in the cancer tissue, while promoting rapid clearance from the body[17]. Finally, they must be traceable throughout the body[18]. Overall, an ideal delivery agent would have the ability to target cancer cells with large numbers of boron-10 atoms, while remaining traceable and nontoxic to the human body. High efficiency boron delivery agent is still under developing. One of the options is using nanoparticles to achieve those points.

Nanoparticles have been used with considerable success in the diagnosis and treatment of cancer[19]. They are often combined with conventional cancer treatment drugs to enable the direct targeting of tumor tissue, thereby reducing the effects of toxicity on the immune system and normal tissue[20]. Nanodiamonds (NDs) provide excellent bio-compatibility[21-24] and their surface is easily modified to facilitate the attachment of conventional drugs that kill tumor cells via chemical processes[25]. Most previous studies in this field have focused exclusively on the surface of NDs (i.e., disregarding the interior of the NDs) for drug delivery or other features. Recently, we developed a novel technique that uses ion implantation to enable the addition of boron-10 to NDs for use as a boron delivery agent. Atoms implanted within NDs are stable and provide good bio-compatibility[26]. This technique allows us to add a new dimension for BNCT applications.

In this work, we fabricate boron-10 embedded nanodiamonds using physical ion implantation in conjunction with a two-step annealing process. During the fabrication process, high-energy boron ions penetrate into the diamond lattice to create vacancies. Annealing is then used to form nitrogen-vacancy (NV) centers within the NDs (first step) and to remove graphite from the surface (second step). It is generally not possible to observe NDs after they enter cells or animal body; however, the proposed boron-10 embedded NDs with red fluorescence can be observed using fluorescence microscopy or an *in vivo* imaging system (IVIS). With optical visibility, boron-10 embedded NDs become an observable boron delivery agent. One has the chance to observe them by red fluorescence with low background signals in this wavelength region[27]. Boron-10 embedded NDs overcome the limitation that the existing boron delivery agents are not easy to be observed inside the cells or animal body.

Furthermore, once combining with targeting ability by NDs surface modification[28], boron-10 embedded NDs can become akin to a kind of Trojan Horse. The Trojan people (cells) will pull the horses into their city (the interior of cells). In other words, boron-10 embedded NDs will more accumulate in tumor and create higher therapeutic efficiency. It will solve the long existing problem that current boron delivery agents have weak targeting performance in the last century. To summarize, fluorescent boron-10 embedded NDs provide a novel boron delivery agent to extend the practical use of NDs and open up new avenues of research for the further development of BNCT.

The NDs powder with average 100 *nm* in diameter from Microdiamant Co. was dissolved in DI water and then the solution was coated on silicon wafer. Boron-10 ions were implanted into NDs with energy of 30 *keV*, dose of $3\times10^{15}$ *ions/cm$^2$* and energy 60 *keV*, dose of $3\times10^{15}$ *ions/cm$^2$*. After implantation, the wafer was cut to many pieces to perform different processes. The annealing processes were performed at 800°C for 2 hours under vacuum and 450°C for 3 hours under air using Nabertherm annealing furnace. X-ray photoelectron spectroscopy(XPS) spectra were obtained using PHI Quantera II system. Microscope fluorescence images were obtained by Nikon Ti-U inverted research fluorescence microscopy. The excitation wavelength range is 560±12.5nm and the detected emission wavelength range is 692±40nm. In vivo imaging system IVIS Lumina II was used to capture IVIS fluorescence images. The excitation wavelength is 535 nm and the detected emission wavelength is 680 nm.

Following boron-10 ion implantation, our first objective was to confirm that the implanted NDs contained boron ions. Fig. 1 presents the B 1s XPS spectra of as-implanted NDs and NDs. The characteristic peak of boron[29] near 190 *eV* can be clearly observed in Fig. 1 (a). The measured spectra revealed that the as-implanted NDs contained boron ions and we called them B-NDs. For practical bio-applications, it is crucial that the nanoparticles used for tracing provide fluorescence of a sufficient intensity to allow observation using a microscope or other device. Many nitrogen atoms are included in the ND structure during fabrication. Boron-10 ion implantation produces many vacancies in the diamond lattice. In this study, we used high-temperature vacuum annealing to promote the formation of NV centers to facilitate red fluorescence. Initial annealing under vacuum at 800°C was followed by air annealing at 450°C for the removal of graphite layers, which might otherwise block the fluorescence. Two-step annealing methods such as this have previously been used to repair the lattice of NDs and remove surface graphite[30]. We used XPS spectra to evaluate the effects of the two-step annealing process on B-NDs.

All obtained XPS spectra were fitted with possible carbon chemical states[31, 32]. Figs. 2 (a)

and (b) respectively present the C 1s spectra of the original (non-implanted) NDs and as-implanted B-NDs. The difference of calculated $sp^3/sp^2$ ratio between the original NDs and as-implanted B-NDs is insignificant. It implies the damage caused by boron-10 ion implantation. Fig. 2 (c) presents the C 1s spectra of B-NDs annealed under vacuum at 800°C. The calculated $sp^3/sp^2$ ratio of the B-NDs increased from 2.74 before annealing to approximately 7 after annealing. This is a clear demonstration of how high-temperature vacuum annealing repairs the lattice structure of as-implanted B-NDs. Fig. 3 (d) presents the C 1s spectrum of B-NDs following two-step annealing. The carbide signal likely originated at the contact surface between the NDs and the silicon substrate. The calculated $sp^3/sp^2$ ratio of the B-NDs increases from 7 prior to second-stage annealing to 11.6 after this step. This clearly demonstrates the efficacy of air annealing at 450°C for the removal of graphite from the surface of the B-NDs. To summarize, first-stage annealing formed NV centers, and second-stage annealing removed surface graphite, thereby allowing the tracking of red fluorescence.

In this study, we employed a microscope commonly used to observe fluorescence in biology labs. Fig. 3 presents fluorescence images of NDs and B-NDs obtained using the microscope under various conditions. Only the B-NDs that underwent the proposed two-step annealing process (Fig. 3 (d)) are clearly observable. Thus, it would be reasonable to expect that even after the B-NDs are internalized by cells, they could be observed in a similar manner. We also used non-invasive IVIS system to verify our findings. The IVIS system uses the full wavelength spectrum from high-energy blue to low-energy infrared to study disease progression, cell trafficking, and gene expression patterns via fluorescence or luminescence[33]. The detection of B-NDs using the IVIS system is a clear indication of the efficacy of these B-NDs in practical bio-applications. The emission wavelength of NV centers of maximum intensity is approximately 680 nm[34]. We used a halogen lamp light source with a wavelength of 535 nm to excite the samples and receive fluorescence emission signals with a wavelength of 680 nm. Fig. 4 presents IVIS fluorescence images of the original NDs and B-NDs under various conditions. The grey image is taken by visible light camera and then the fluorescent signal is overlaid on this grey image. Only the colored area presented fluorescence emissions at 680 nm. Again, the only visible fluorescence was from B-NDs after two-step annealing. Testing was also conducted on NDs that underwent the proposed two-step annealing process without boron-10 implantation. Those NDs remained invisible under the instruments used in this study. Visibility under a fluorescence microscope and the IVIS system provides compelling evidence of the practical value of the proposed B-NDs.

In conclusion, this paper reports the fabrication of fluorescent boron-10 embedded NDs using boron-10 ion implantation followed by a two-step annealing process. The B-NDs annealed at 800°C for 2 hours under vacuum and 450°C for 3 hours under air provided red fluorescence of sufficient intensity to enable detection using a commercial fluorescence microscope as well as the IVIS system. The proposed B-NDs, combining optical visibility and boron-10 transport capability, are a promising boron delivery agent suitable for a wide range of biomedical applications.

**Acknowledgement**

The authors acknowledge Team Union Ltd. for their financial supporting. The authors thank Dr. Causon K.C. Jen at Axcelis Technologies., Inc. for his valuable inputs.

**Figure caption**

Fig. 1. B 1s XPS spectra of (a) as-implanted NDs. (b) original NDs.

Fig. 2. C 1s XPS spectra of (a) Original NDs. (b) as-implanted B-NDs. (c) B-NDs annealed at 800°C under vacuum. (d) B-NDs annealed at 800°C under vacuum and 450°C in air.

Fig. 3. Microscope fluorescence images of NDs and B-NDs under different condition. (a) Original NDs. (b) As implanted B-NDs (c) B-NDs annealed at 800°C under vacuum. (d) B-NDs annealed at 800°C under vacuum and 450°C in air. Scale bar: 10 μm.

Fig. 4. IVIS fluorescence images of NDs and B-NDs under different condition. (a) Original NDs. (b) As implanted B-NDs (c) B-NDs annealed at 800°C under vacuum. (d) B-NDs annealed at 800°C under vacuum and 450°C in air. The grey image is taken by visible light camera and then the fluorescent signal is overlaid on this grey image. Only the colored area presented fluorescence emissions at 680 nm.

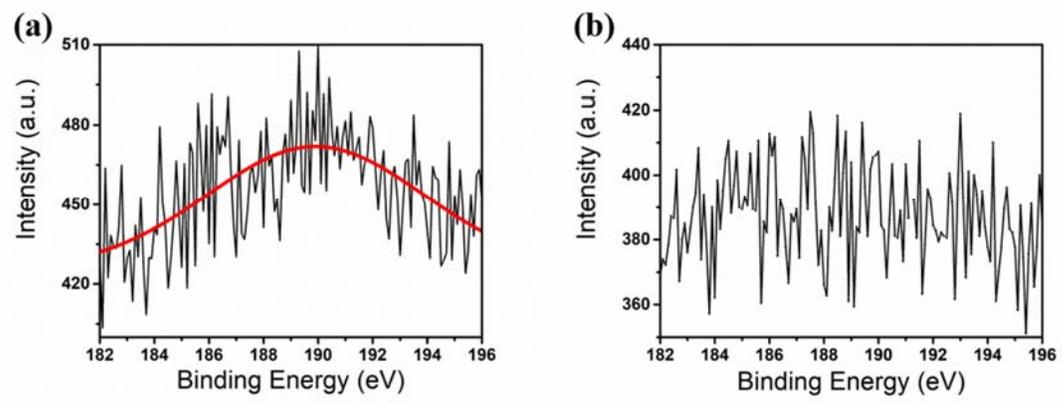

Fig. 1.

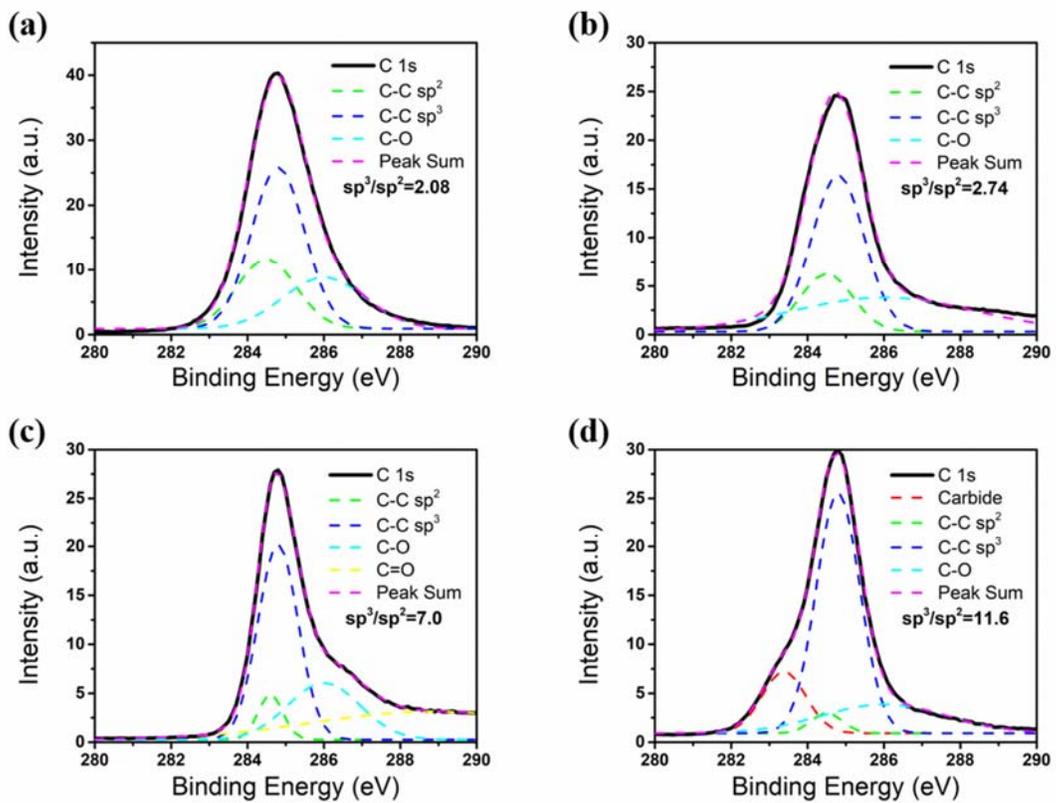

Fig. 2.

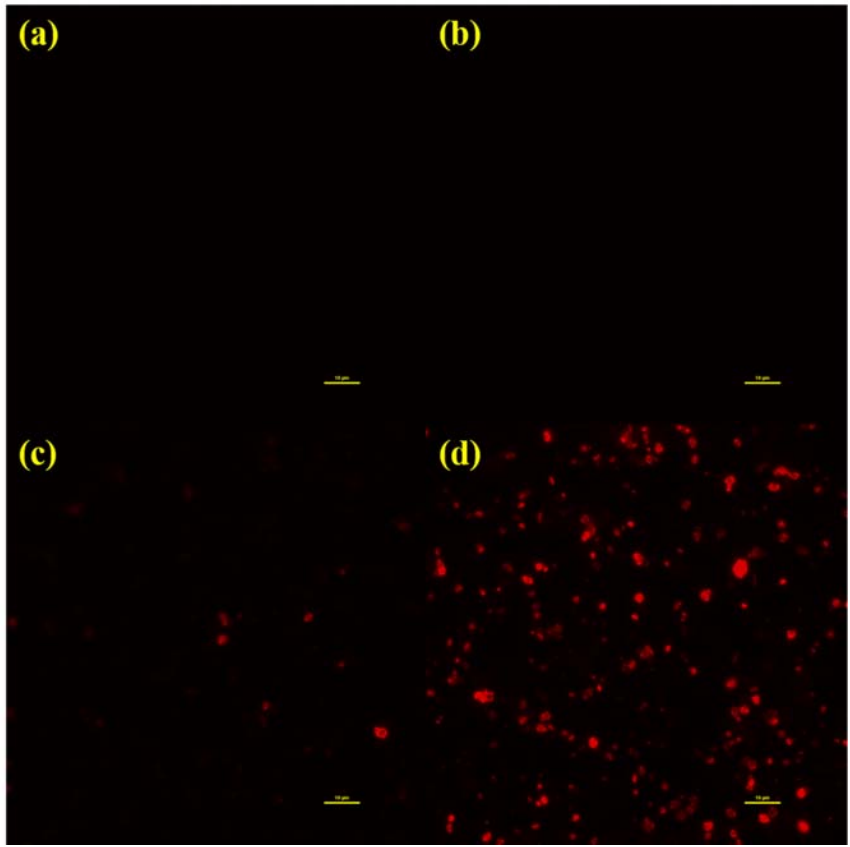

Fig. 3.

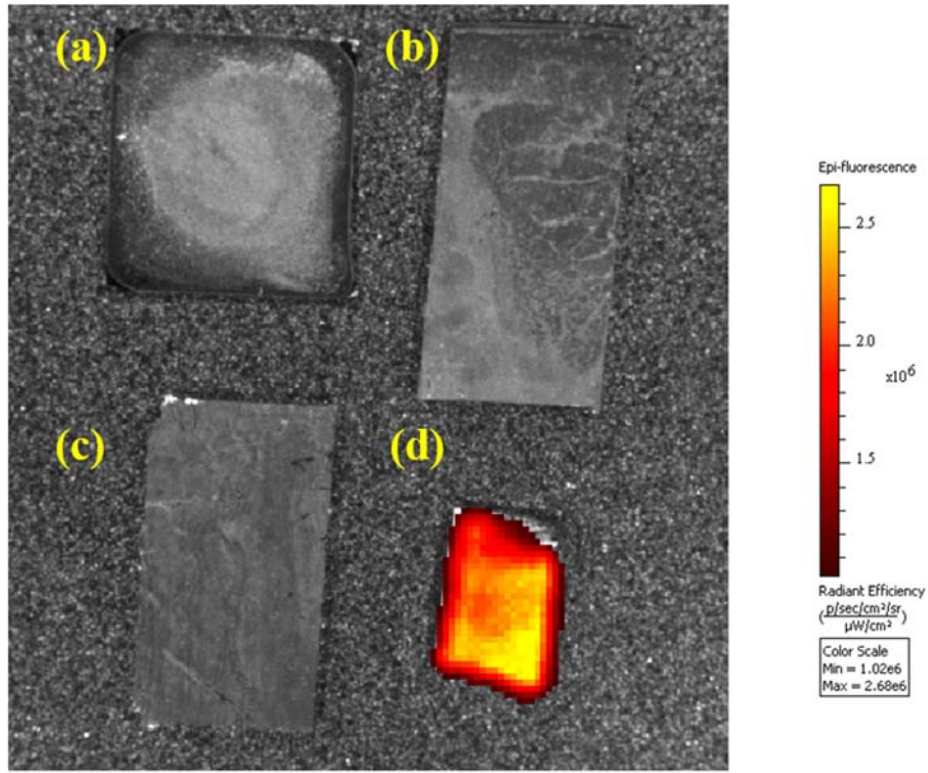

Fig. 4.